\documentclass[prd,%
preprint,%
tightenlines,showpacs,%
nofootinbib,%
superscriptaddress,%
amsfonts,amsmath]{revtex4}

\begin{document}

\title{Time-dependent spherically symmetric covariant Galileons}

\author{Eugeny~Babichev} \email{eugeny.babichev@th.u-psud.fr}
\affiliation{Laboratoire de Physique Th\'eorique d'Orsay,
B\^atiment 210, Universit\'e Paris-Sud 11, F-91405 Orsay Cedex,
France}
\affiliation{UPMC-CNRS, UMR7095,
Institut d'Astrophysique de Paris,
${\mathcal{G}}{\mathbb{R}}\varepsilon{\mathbb{C}}{\mathcal{O}}$,
98bis boulevard Arago, F-75014 Paris, France}

\author{Gilles~\surname{Esposito-Far\`ese}} \email{gef@iap.fr}
\affiliation{UPMC-CNRS, UMR7095,
Institut d'Astrophysique de Paris,
${\mathcal{G}}{\mathbb{R}}\varepsilon{\mathbb{C}}{\mathcal{O}}$,
98bis boulevard Arago, F-75014 Paris, France}

\begin{abstract}
We study spherically symmetric solutions of the cubic covariant
Galileon model in curved spacetime in presence of a matter
source, in the test scalar field approximation. We show that a
cosmological time evolution of the Galileon field gives rise to
an {\it induced} matter-scalar coupling, due to the
Galileon-graviton kinetic braiding, therefore the solution for
the Galileon field is non trivial even if the bare matter-scalar
coupling constant is set to zero. The local solution crucially
depends on the asymptotic boundary conditions, and in particular,
Minkowski and de Sitter asymptotics correspond to different
branches of the solution. We study the stability of these
solutions, namely, the well-posedness of the Cauchy problem and
the positivity of energy for scalar and tensor perturbations, by
diagonalizing the kinetic terms of the spin-2 and spin-0 degrees
of freedom. In addition, we find that in presence of a
cosmological time evolution of the Galileon field, its kinetic
mixing with the graviton leads to a friction force, resulting to
efficient damping of scalar perturbations within matter.
\end{abstract}

\date{December 6, 2012}


\maketitle

\section{Introduction}
\label{Sec1}
A scalar-tensor theory of gravity with higher-order derivative
terms in the action, but with no more than second derivatives in
the field equations, has first been found in
Ref.~\cite{Horndeski}. Later, a subclass of the scalar part of
this theory has been discussed in a different context
in~\cite{Fairlie}. In the framework of the
Dvali-Gabadadze-Poratti (DGP) brane model~\cite{Dvali:2000hr},
the cubic scalar higher derivative action appears naturally in a
specific decoupling limit. More recently, a generalization of the
DGP decoupling limit action has been presented in
\cite{Nicolis:2008in}, where this model was dubbed ``Galileon'',
because the considered action is invariant under the galilean
transformation of the field (adding a constant to the field or a
constant vector to its gradient). The original Galileon model has
been later covariantized to include a dynamical metric
\cite{Deffayet:2009wt}, and then extended and generalized in
several papers, see, e.g. \cite{Deffayet:2009mn,Deffayet:2011gz}.

The fact that the field equations are no more than second order,
in spite of the higher-order action, guarantees the absence of
any Ostrogradsky ghost, i.e., an extra degree of freedom with
negative kinetic energy. Another interesting feature of the
Galileon model is that it generically possesses a Vainshtein
screening mechanism. This last property can be understood as a
particular case of ``k-mouflage'', i.e., of kinetic
screening~\cite{Babichev:2009ee}. The Vainshtein mechanism hides
the propagating scalar degree of freedom, thus providing a way to
pass local gravity tests, even if the scalar field is directly
coupled to matter. The local effects of the Vainshtein screening
of the Galileon has been studied in a number of
works~\cite{galileons}.

Usually, when studying local effects of the Galileon, the
cosmological evolution of the scalar field is not taken into
account, because of the smallness of the time derivative of the
field compared to the spacial gradients. However, such a
disregard of the Galileon's time evolution may be misleading. For
example, it was shown in Ref.~\cite{Babichev:2011iz} that in
spite of the Vainshtein mechanism, which operates locally, this
time evolution still poses severe constraints on many interesting
scalar-tensor theories possessing a shift symmetry and a
matter-scalar coupling.

In this work, we study the behavior of the covariant cubic
Galileon model (the covariant version of the decoupling limit of
the DGP model) in presence of a spherically symmetric matter
source, and possible time evolution of the scalar field imposed
by cosmology. We investigate in detail two distinct cases: (i)~a
spherically symmetric matter source is embedded in asymptotically
Minkowski spacetime, and the Galileon field does not depend on
time; (ii)~asymptotically de Sitter Universe with time-dependent
Galileon field. To realize the latter regime, we adopt the
cosmological solution of the Kinetically Gravity Braiding
model~\cite{Deffayet:2010qz}, for which the acceleration of the
Universe is driven by the time evolving Galileon field itself. We
find non-homogeneous solutions in the test scalar field
approximation, and analyze the stability of perturbations on top
of these solutions.

The paper is organized as follows. In Sec.~\ref{Sec2}, we
introduce the action and field equations of the cubic covariant
Galileon. In Sec.~\ref{Sec3}, we derive a solution for a scalar
field having a cosmological time evolution, in an arbitrary
spherically symmetric static spacetime, in the test field
approximation. Perturbations are investigated in Sec.~\ref{Sec4},
where we diagonalize the kinetic matrix, demixing the spin-0 and
spin-2 propagating modes. The solution in asymptotically
Minkowski spacetime, in presence of a matter source with no time
evolution of the scalar field, is studied in Sec.~\ref{Sec5},
where we find the effective metric for the helicity-0 propagating
mode, and the conditions for stability of the solution. In
Sec.~\ref{Sec6}, we perform a similar analysis for asymptotically
de Sitter spacetime, in presence of a cosmological evolution of
the scalar field. Scalar perturbations inside a matter source are
investigated in Sec.~\ref{Sec7}. Besides the analyses of the
Cauchy problem and the positivity of energy, we also exhibit a
large friction term, which is caused by the
cosmologically-imposed time evolution of the Galileon within
matter. Our conclusions are presented in Sec.~\ref{Concl}.

\section{Action and field equation}
\label{Sec2}
We consider the quadratic plus cubic Galileon action in curved
spacetime,
\begin{equation}
S = 2 M_P^2 \int{d^4 x\, \sqrt{-g}\left\{\frac{R}{4}
- \frac{k_2}{2}(\partial_\mu\varphi)^2
-\frac{k_3}{2M^2} \Box\varphi (\partial_\mu\varphi)^2\right\}}
+S_\text{matter}\left[\psi_\text{matter};
\tilde g_{\mu\nu}\right],
\label{Eq1}
\end{equation}
where the scalar field $\varphi$ is chosen to be dimensionless,
where the reduced Planck mass\footnote{Throughout this paper, we
choose natural units such that $\hbar = c = 1$.} $M_P \equiv
(8\pi G)^{-1/2}$ should not be confused with the mass scale $M$
entering the Galileon's cubic kinetic term, and where $g$ and $R$
denote respectively the determinant and the scalar curvature of
the metric $g_{\mu\nu}$ (of signature $-+++$) used to define
covariant derivatives and to contract indices. All matter fields,
globally denoted as $\psi_\text{matter}$, are assumed to be
minimally coupled to a physical (``Jordan'') metric $\tilde
g_{\mu\nu} \equiv e^{2\alpha\varphi}g_{\mu\nu}$, where $\alpha$
denotes a dimensionless matter-scalar coupling constant. The
constants $k_2$ and $k_3$, dimensionless too, could \textit{a
priori} be reabsorbed in the definitions of the scalar field
$\varphi$ and of the mass scale $M$ (while changing
simultaneously the value of $\alpha$:
$\varphi^\text{new}= |k_2|^{1/2} \varphi$,
$M^\text{new}= M |k_2|^{3/4}/|k_3|^{1/2}$,
$\alpha^\text{new}= \alpha/|k_2|^{1/2}$). In other words, the
model (\ref{Eq1}) actually depends on only two independent
parameters, for instance $M$ and $\alpha$. However, the redundant
factors $k_2$ and $k_3$ will be useful in the following to keep
track of the origin of the different terms, and above all to
change easily the signs of these two kinetic terms. Their
numerical factors are chosen to simplify the following results
involving $k_3$, and so that $k_2 = 1$, $k_3 = 0$ defines a
canonically normalized positive-energy spin-0 degree of freedom.
Another interest of these coefficients is that any normalization
convention can easily be recovered, for instance $k_2 =
\frac{1}{2}$ corresponding to the relative weight of the
Einstein-Hilbert action and the scalar-field standard
kinetic term in many cosmological papers.

When $k_3 = 0$, i.e., in standard scalar-tensor theories of
gravity \cite{Jordan,Willbook,Damour:1992we},
$g_{\mu\nu}$ is called the ``Einstein metric'' and its
fluctuations describe a spin-2 degree of freedom. As we will see
in Sec.~\ref{Sec4}, this is no longer the case when $k_3\neq 0$
because of the metric-scalar derivative coupling entering the
Galileon's cubic kinetic term. The metric $g_{\mu\nu}$ is thus
neither the Jordan ($\tilde g_{\mu\nu}$) nor the Einstein one,
and it should just be considered as the variable we choose to
write the field equations as simply as possible.

We shall study a \textit{test} scalar field in the background
metric generated by a spherical body, assuming that its
backreaction on this metric is negligible, which is always the
case for a small enough matter-scalar coupling
constant\footnote{Anticipating on definition (\ref{Eq10}) below,
we are actually talking here about the \textit{effective}
matter-scalar coupling constant $\alpha_\text{eff}$, which takes
into account the induced coupling in presence of a cosmological
time evolution of the scalar field.}. We shall also check
\textit{a posteriori} in which conditions this backreaction is
indeed negligible. It will thus suffice to focus first on the
scalar field equation, which can be written as
\begin{equation}
\nabla_\mu J^\mu = -\alpha T,
\label{Eq2}
\end{equation}
where $J^\mu \equiv -(1/\sqrt{-g})\delta S/\delta
\partial_\mu\varphi$ is the scalar field's current, and $T$
denotes the trace of $T^{\mu\nu} \equiv (2/\sqrt{-g}) (\delta
S_\text{matter}/\delta g_{\mu\nu})$, related to the physical
(Jordan-frame) matter energy-momentum tensor
$\tilde T^{\mu\nu} \equiv (2/\sqrt{-\tilde g})
(\delta S_\text{matter}/\delta \tilde g_{\mu\nu})$ by
$T^{\mu\nu} = e^{6\alpha\varphi}\, \tilde T^{\mu\nu}
\Rightarrow T = e^{4\alpha \varphi}\, \tilde T$. The
current $J^\mu$ reads explicitly
\begin{equation}
\frac{1}{M_P^2} J^\mu = 2 k_2 \partial^\mu\varphi
+ 2 \frac{k_3}{M^2} \Box\varphi \partial^\mu\varphi
- \frac{k_3}{M^2} \nabla^\mu
\left((\partial_\lambda\varphi)^2\right),
\label{Eq3}
\end{equation}
and Eq.~(\ref{Eq2}) takes thus the full form
\begin{equation}
k_2\Box\varphi + \frac{k_3}{M^2}
\left\{(\Box\varphi)^2-(\nabla_\mu\partial_\nu\varphi)^2
-R^{\mu\nu}\partial_\mu\varphi\partial_\nu\varphi\right\}
= -\frac{\alpha}{2M_P^2} T,
\label{Eq4}
\end{equation}
where the Ricci tensor $R^{\mu\nu}$ enters because of a
difference of third-order covariant derivatives of the scalar
field:
$[\nabla_\mu\nabla_\nu-\nabla_\nu\nabla_\mu]\nabla^\nu\varphi =
-R_\mu^\nu \partial_\nu \varphi$. Note that this curvature tensor
involves second derivatives of the metric, therefore
Eq.~(\ref{Eq4}) actually mixes spin-0 and spin-2 degrees of
freedom. We will diagonalize them when studying perturbations in
Sec.~\ref{Sec4}, but to derive the spherically-symmetric solution
for a test scalar field, it suffices to replace this Ricci tensor
by its vacuum value $R^{\mu\nu} = 0$ outside the central massive
body.

\section{Background solution}
\label{Sec3}
A static and spherically symmetric background metric reads in
Schwarzschild coordinates
\begin{equation}
ds^2 = - e^{\nu(r)} dt^2 + e^{\lambda(r)}dr^2 + r^2 d\Omega^2,
\label{Eq5}
\end{equation}
where $\lambda(r)$ and $\nu(r)$ are two functions of the radial
coordinate. The exterior (vacuum) solution is well known to read
$e^\nu = e^{-\lambda} = 1-r_S/r$ in an asymptotically flat
spacetime, where $r_S \equiv 2Gm$ denotes the Schwarzschild
radius of the central body of mass $m$, but we will also consider
the Schwarzschild-de Sitter solution in Sec.~\ref{Sec6} below.
We look for a test scalar field solution of Eq.~(\ref{Eq2}) or
(\ref{Eq4}), assuming that the cosmological evolution imposes a
linear time dependence, i.e., that $\ddot \varphi = 0$ (where a
dot denotes a time derivative). As shown in \cite{Deffayet:2010qz},
such a linear time-evolution is a cosmological attractor. Let us
look for a solution of the form
\begin{equation}
\varphi = \phi(r) + \dot\varphi_c t +\varphi_0,
\label{Eq6}
\end{equation}
where $\dot\varphi_c$ and $\varphi_0$ are assumed to be
constants. The ansatz (\ref{Eq6}) has been used in a similar
context for the study of the Galileon accretion
\cite{Babichev:2010kj} and the effects of cosmologically evolving
scalar field on the variation of the Newton constant
\cite{Babichev:2011iz}. Actually, since $\varphi' = \phi'$ (where
a prime denotes a radial derivative), the notation $\phi$ will
not be useful in the following, and we can thus write
Eq.~(\ref{Eq2}) in terms of $\varphi'$. With the ansatz
(\ref{Eq6}), this field equation reduces in vacuum to a mere
$\partial_r \left[r^2 e^{(\lambda+\nu)/2}J^r\right] = 0$, and it
can be integrated once as $e^{(\lambda+\nu)/2} J^r = \alpha M_P^2
r_S/r^2$, the constant of integration being imposed by the same
Eq.~(\ref{Eq2}) within matter.\footnote{\label{foot2}Actually,
this constant of integration is slightly modified by self-gravity
effects within the body, because the scalar field is sourced by
the \textit{trace} of the matter energy-momentum tensor, with a
different pressure dependence than in Einstein's equations. We
should thus write rigorously, like in Brans-Dicke theory
\cite{Willbook,Damour:1992we},
$e^{(\lambda+\nu)/2} J^r = \alpha (1-2s) M_P^2
r_S/r^2$, where $s \equiv -\partial\ln m/ \partial\ln G \sim
|E_\text{grav}|/m \sim r_S/r_\text{body}$.} In terms of
$\varphi'$, this reads
\begin{equation}
\frac{2 k_3}{M^2r}\left(1+\frac{\nu'r}{4}\right)e^{-\lambda}
\varphi'^2 + k_2 \varphi'
-\frac{e^{(\lambda-\nu)/2}}{2}\left(
\frac{\alpha r_S}{r^2} + \frac{k_3}{M^2}\,\dot\varphi_c^2\,
\frac{\nu'}{e^{(\lambda+\nu)/2}} \right) = 0,
\label{Eq7}
\end{equation}
whose two solutions are
\begin{equation}
\varphi' = -\frac{k_2 M^2 e^\lambda r}{4 k_3\left(1
+\frac{\nu' r}{4}\right)}
\left[1\pm\sqrt{1+\frac{4 k_3 r_S}{k_2^2
M^2 r^3 e^{(\lambda+\nu)/2}}
\left(1+\frac{\nu' r}{4}\right)
\left(\alpha + \frac{k_3}{M^2}\,\dot\varphi_c^2\,
\frac{\nu' r^2}{r_S e^{(\lambda+\nu)/2}}\right)}\right].
\label{Eq8}
\end{equation}
Note that for a test scalar field, whose backreaction on the
metric is negligible, these expressions (\ref{Eq8}) are exact,
i.e., correct to any post-Newtonian order (while taking into
account the slight numerical change of $\alpha \rightarrow \alpha
(1-2s)$ due to self-gravity, as underlined in
footnote~\ref{foot2}). Of course, for the exterior Schwarzschild
solution, they can be simplified because $e^{\lambda+\nu} = 1$.
Moreover, for $r \gg r_S$, one may expand $\lambda$ and $\nu$ in
powers of $r_S/r$, and get
\begin{equation}
\varphi' = -\frac{k_2 M^2 r}{4 k_3}
\left(1\pm\sqrt{1+\frac{4 k_3 r_S}{k_2^2 M^2 r^3}\,
\alpha_\text{eff}}\,\right)\times
\left[1+\mathcal{O}\left(\frac{r_S}{r}\right)\right],
\label{Eq9}
\end{equation}
where
\begin{equation}
\alpha_\text{eff} \equiv \alpha
+ \frac{k_3 \dot\varphi_c^2}{M^2}
\label{Eq10}
\end{equation}
is an effective matter-scalar coupling constant, modifying the
bare one, $\alpha$, because of the self-interaction of the scalar
field with its own cosmologically-imposed time evolution
$\dot\varphi_c$. Let us underline that even if one assumes
$\alpha = 0$ in action (\ref{Eq1}), i.e., \textit{a priori} no
matter-scalar interaction, the cosmological time evolution of the
scalar field does generate a nonvanishing coupling to matter.
This is due to the quadratic terms entering the current
(\ref{Eq3}), since the Christoffel symbols do depend on the
background metric and thereby on the mass of the central body.

We will discuss in Secs.~\ref{Sec5} and \ref{Sec6} which sign
should be chosen in solution (\ref{Eq8}), by studying the
large-distance behavior of the scalar field. We will see in
particular that it does not reduce to (\ref{Eq9}) at
cosmologically large distances if spacetime is not assumed to be
asymptotically flat (see Eqs.~(\ref{Eq30}) to (\ref{Eq32})). At
present, let us just note that the square root entering
(\ref{Eq9}) involves a contribution $\propto 1/r^3$, dominating
at small enough distances (still assumed to be much larger than
the Schwarzschild radius $r_S$). We define the ``Vainshtein
radius'' as
\begin{equation}
r_V \equiv \left(\frac{4 k_3 r_S}{M^2 k_2^2}
\alpha_\text{eff}\right)^{1/3}.
\label{Eq11}
\end{equation}
It is positive only if $k_3 \alpha_\text{eff} > 0$, which is
always the case when the bare $\alpha = 0$, but depends on the
model when $\alpha \neq 0$. When $r_V < 0$, then Eqs.~(\ref{Eq8})
or (\ref{Eq9}) simply do not give an acceptable solution for the
scalar field at radii $r < |r_V|$. This means that the ansatz
(\ref{Eq6}) cannot be correct at such small distances, and the
actual solution must mix its time and radial dependences in a
more subtle way. When $r_V$ is positive, on the other hand,
Eq.~(\ref{Eq8}) or (\ref{Eq9}) is a correct solution. In the
range $r_S \ll r \ll r_V$, the scalar field then takes the form
\begin{subequations}
\label{Eq12}
\begin{eqnarray}
\varphi' &=& \mp\frac{\text{sign}(k_2 k_3)}{2}\,
\sqrt{\frac{r_S}{r}}\sqrt{\frac{M^2\alpha_\text{eff}}{k_3}}
\left[1+\mathcal{O}\left(\frac{r_S}{r}\right)
+\mathcal{O}\left(\frac{r^3}{r_V^3}\right)\right]
\label{Eq12a}\\
&=& \mp\frac{k_2 M^2}{4 k_3}
\sqrt{\frac{r_V^3}{r}}
\left[1+\mathcal{O}\left(\frac{r_S}{r}\right)
+\mathcal{O}\left(\frac{r^3}{r_V^3}\right)\right].
\label{Eq12b}
\end{eqnarray}
\end{subequations}
Note in passing that the product $k_3\varphi'$, which will play
an important role in the following sections, has a sign which
does not depend on $k_3$ (provided this Vainshtein regime exists,
i.e., notably that $k_3 \alpha_\text{eff} > 0$), but which does
depend on the sign of $k_2$, on the contrary.

It should be underlined that this regime $r_S \ll r \ll r_V$ can
exist even when $\alpha = 0$, provided
$|2k_3\dot\varphi_c/k_2M^2| \gg r_S$. Therefore, although one
could have naively deduced from (\ref{Eq8})-(\ref{Eq9}) that
$\varphi'$ either vanishes or is proportional to $r$ (depending
on the sign to be taken in this solution), Eq.~(\ref{Eq12})
actually proves it is proportional to $r^{-1/2}$. The homogeneous
scalar field assumed in Ref.~\cite{Deffayet:2010qz} thus cannot
be used in the vicinity of a massive body, when there is a
nonzero cosmological time evolution $\dot\varphi_c \neq 0$.
It would remain possible to assume that $\alpha_\text{eff} = 0$,
but this would need to fine-tune the cosmologically-driven time
derivative $\dot\varphi_c$ to the precise value
$(-\alpha/k_3)^{1/2}M$ involving the constants parameters of
action (\ref{Eq1}) (while also assuming that $\alpha k_3 < 0$).

\section{Perturbations}
\label{Sec4}
Let us now study the perturbations of both the metric and the
scalar field around the background solution defined by
Eqs.~(\ref{Eq5}) and (\ref{Eq8}). We write
$g_{\mu\nu}^\text{full} = g_{\mu\nu} + h_{\mu\nu}$ and
$\varphi^\text{full} = \varphi + \pi$, where $g_{\mu\nu}$ and
$\varphi$ denote the background fields, and $h_{\mu\nu}$ and
$\pi$ their perturbations. Our aim is to identify the pure
helicity-2 and 0 degrees of freedom, to analyze in which
conditions they both carry positive energy, so that any ghost
instability is avoided. We expand action (\ref{Eq1}) to second
order in the perturbations, and keep only the kinetic terms,
which involve at least two derivatives of these fields. Of
course, terms of the form $f(\text{background}) h \nabla\nabla
\pi$ can be integrated by parts, and contribute thus to the
kinetic terms as $-f(\text{background}) \nabla h \nabla \pi$.
Because of the cubic Galileon interaction in action (\ref{Eq1}),
one actually also gets a term \hbox{$\propto 2\partial_\mu\varphi
\partial^\mu\pi \Box\pi$}, involving \textit{three} derivatives
of the scalar perturbation $\pi$, but this can be rewritten as
$\Box\varphi(\partial_\mu\pi)^2 - 2 \nabla_\mu\partial_\nu\varphi
\partial^\mu\pi \partial^\nu\pi$ by partial integration, as
expected because the Galileon field equations (perturbed or not)
are known to involve at most second derivatives. The final sum of
all kinetic terms for the perturbations $h_{\mu\nu}$ and $\pi$
reads
\begin{eqnarray}
\frac{1}{2M_P^2}\,\frac{\mathcal{L}_2^\text{kinetic}}{\sqrt{-g}}
&=& -\frac{1}{8} \nabla_\mu
h_{\alpha\beta}P^{\alpha\beta\gamma\delta}
\nabla^\mu h_{\gamma\delta}
+\frac{1}{8}\left(h^\lambda_{\nu;\lambda}
-\frac{1}{2}h_{,\nu}\right)^2
-\frac{k_2}{2}\left(\partial_\mu\pi\right)^2
\nonumber\\
&&-\frac{k_3}{2M^2}\biggl[
2\Box\varphi \left(\partial_\mu\pi\right)^2
- 2 \nabla_\mu\partial_\nu\varphi\,
\partial^\mu\pi \partial^\nu\pi
+ \partial_\mu\varphi\partial_\nu\varphi\,
\partial_\lambda\pi\nabla^\lambda h^{\mu\nu}
\nonumber\\
&&\hphantom{-\frac{k_3}{2M^2}}
- 2\partial^\mu\varphi \partial^\nu\varphi\,
\partial_\mu\pi\, \left(h^\lambda_{\nu;\lambda}
-\frac{1}{2}h_{,\nu}\right)
\biggr],
\label{Eq13}
\end{eqnarray}
where $P^{\alpha\beta\gamma\delta}\equiv
\frac{1}{2}g^{\alpha\gamma}
g^{\beta\delta}-\frac{1}{4}g^{\alpha\beta}g^{\gamma\delta}$, the
indices of $h_{\mu\nu}$ are raised with the background metric
$g^{\rho\sigma}$, and $h\equiv h^\lambda_\lambda$ is the trace of
$h_{\mu\nu}$. The presence of cross terms $f(\text{background})
\nabla\pi \nabla h$ illustrates that the helicity-2 and 0 degrees
of freedom are mixed. But a simple change of variables actually
suffices to diagonalize these kinetic terms. We define
\begin{equation}
\hbar_{\mu\nu} \equiv h_{\mu\nu}
+\frac{4 k_3}{M^2}\left[\partial_\mu\varphi\partial_\nu\varphi
-\frac{1}{2}g_{\mu\nu}
\left(\partial_\lambda\varphi\right)^2\right] \pi,
\label{Eq14}
\end{equation}
and Eq.~(\ref{Eq13}) then takes the form
\begin{equation}
\frac{1}{2 M_P^2}\,\frac{\mathcal{L}_2^\text{kinetic}}{\sqrt{-g}}
= -\frac{1}{8} \nabla_\mu \hbar_{\alpha\beta}
P^{\alpha\beta\gamma\delta}
\nabla^\mu \hbar_{\gamma\delta}
+\frac{1}{8}\left(\hbar^\lambda_{\nu;\lambda}
-\frac{1}{2}\hbar_{,\nu}\right)^2
-\frac{1}{2}\mathcal{G}^{\mu\nu}\partial_\mu\pi\partial_\nu\pi,
\label{Eq15}
\end{equation}
where
\begin{equation}
\mathcal{G}^{\mu\nu} \equiv
g^{\mu\nu} \left[k_2 + \frac{2k_3}{M^2}\Box\varphi
- \frac{k_3^2}{M^4}\left(\partial_\lambda\varphi\right)^4 \right]
-\frac{2k_3}{M^2}\nabla^\mu\partial^\nu\varphi
+4 \frac{k_3^2}{M^4}\left(\partial_\lambda\varphi\right)^2
\partial^\mu\varphi\partial^\nu\varphi,
\label{Eq16}
\end{equation}
proving that $\hbar_{\mu\nu}$ describes a pure helicity-2 field
propagating in the curved background of $g_{\mu\nu}$, and that
the pure helicity-0 field $\pi$ propagates in the effective
metric $\mathcal{G}^{\mu\nu}$. Our full diagonalization recovers
thus this effective metric first derived in
\cite{Deffayet:2010qz}, from a triangulation of the kinetic terms
(using the clever trick of replacing the Ricci tensor entering
Eq.~(\ref{Eq4}) by its source, obtained from the Einstein
equations). The $\mathcal{O}(k_3^2)$ terms are the only
subtleties introduced by this diagonalization, but we will see in
the following that they are subdominant in natural situations.
The crucial information brought by Eq.~(\ref{Eq15}) is that the
spin-2 graviton is never a ghost, and that the effective metric
(\ref{Eq16}) should be of signature $-+++$ for the scalar
perturbation $\pi$ to carry positive energy and have a well-posed
Cauchy problem. We shall perform this analysis in
Secs.~\ref{Sec5} and \ref{Sec6}, for two different asymptotic
conditions.

Before entering this discussion, let us underline that our
diagonalization (\ref{Eq15}) also allows us to exhibit the
induced matter-scalar coupling already noticed in
Eq.~(\ref{Eq10}) for the background solution. We can show now
that the scalar perturbation $\pi$ is also directly coupled to
matter, even when the bare coupling constant $\alpha$ vanishes.
Indeed, since matter is assumed to be minimally coupled to
$\tilde g^\text{full}_{\mu\nu}
= \text{exp}(2\alpha\varphi^\text{full}) g^\text{full}_{\mu\nu}$
(where ``full'' means as before the background fields plus their
perturbations), the action of a point particle reads
\begin{eqnarray}
S_\text{matter} &=& -\int mc
\left(\tilde g^\text{full}_{\mu\nu} dx^\mu dx^\nu\right)^{1/2}
\nonumber\\
&=& -\int mc\, e^{\alpha \varphi}
\left[1+\alpha\pi+\mathcal{O}(\pi^2)\right]
\left(g_{\mu\nu} dx^\mu dx^\nu\right)^{1/2}
\left[1-\frac{1}{2}h_{\rho\sigma}
u^\rho u^\sigma+\mathcal{O}(h^2)\right],
\label{Eq17}
\end{eqnarray}
where $u^\lambda \equiv dx^\lambda/ (g_{\mu\nu} dx^\mu
dx^\nu)^{1/2}$ denotes the unit 4-velocity of the particle (with
respect to the background metric $g_{\mu\nu}$). Although there
does not seem to exist any linear coupling to $\pi$ when $\alpha
= 0$, we actually know that $h_{\mu\nu}$ describes a mixing of
spin-2 and spin-0 degrees of freedom. If we replace it in terms
of $\pi$ and the actual spin-2 excitation $\hbar_{\mu\nu}$,
Eq.~(\ref{Eq14}), we thus get
\begin{eqnarray}
S_\text{matter} = -\int mc\, \left(\tilde g_{\mu\nu}
dx^\mu dx^\nu\right)^{1/2}
\biggl[1&+&\left\{\alpha
+\frac{2k_3}{M^2}\left(u^\lambda\partial_\lambda\varphi\right)^2
+\frac{k_3}{M^2}\left(\partial_\lambda\varphi\right)^2
\right\}\pi
\nonumber\\
&-&\frac{1}{2}\hbar_{\rho\sigma}u^\rho u^\sigma
+\mathcal{O}\left((\hbar_{\rho\sigma},\pi)^2\right)\biggr].
\label{Eq18}
\end{eqnarray}
The quantity within curly brackets plays the role of an effective
linear coupling constant of matter to the scalar perturbation
$\pi$, and it can be nonzero even if $\alpha = 0$. It should
\textit{a priori} not be confused with $\alpha_\text{eff}$,
defined in Eq.~(\ref{Eq10}), which described the effective
coupling of the \textit{background} scalar field $\varphi$ to the
matter source. However, at lowest post-Newtonian order, they
actually coincide when they are constant, notably in a background
such that $\dot\varphi_c\neq 0$ but $\partial_r\varphi = 0$.
Indeed, we have $-\left(\partial_\lambda\varphi\right)^2 =
e^{-\nu}\dot\varphi_c^2 =
\left(u^\lambda\partial_\lambda\varphi\right)^2 +\mathcal{O}(v)$,
where $v$ is the particle's velocity, so that the quantity within
curly brackets in (\ref{Eq18}) reduces to (\ref{Eq10}).

\section{Asymptotic Minkowski spacetime}
\label{Sec5}
We consider in this section that spacetime is asymptotically
Minkowskian, i.e., we impose that the spherically symmetric
metric (\ref{Eq5}) is given by the Schwarzschild solution $e^\nu
= e^{-\lambda} = 1-r_S/r$, and we consistently assume that the
background scalar field has no cosmological time evolution,
$\dot\varphi_c = 0$. Then the upper sign of solution (\ref{Eq8}),
(\ref{Eq9}) or (\ref{Eq12}) must be discarded, otherwise the
scalar field would diverge at spatial infinity (as well as its
derivative and its energy-momentum tensor). With the lower
sign, Eq.~(\ref{Eq9}) gives for $r\rightarrow\infty$ the standard
behavior of a Brans-Dicke scalar field,
\begin{subequations}
\label{Eq19}
\begin{eqnarray}
\varphi' &=& \frac{\alpha r_S}{2 k_2 r^2}
\left[1+\mathcal{O}\left(\frac{r_S}{r}\right)
+\mathcal{O}\left(\frac{r_V^3}{r^3}\right)\right]
\label{Eq19a}\\
\Rightarrow\quad\varphi &=&
\varphi_0 - \frac{\alpha r_S}{2 k_2 r}
\left[1+\mathcal{O}\left(\frac{r_S}{r}\right)
+\mathcal{O}\left(\frac{r_V^3}{r^3}\right)\right].
\label{Eq19b}
\end{eqnarray}
\end{subequations}
In this large-distance regime, the perturbations of the scalar
field also behave as in Brans-Dicke theory, i.e., they propagate
in an effective metric (\ref{Eq16}) dominated by its standard
$k_2g^{\mu\nu}$ contribution. In conclusion, $k_2$ \textit{must}
be positive for the scalar degree of freedom to carry positive
energy in the asymptotic $r\rightarrow\infty$ region.

On the other hand, in the Vainshtein regime $r\ll r_V$ (which
exists only if $k_3\alpha > 0$), the background scalar field
takes the form (\ref{Eq12}) with its lower sign, and the
effective metric (\ref{Eq16}), in which scalar perturbations
propagate, is dominated by its $k_3$ contributions. Indeed,
we have $(2k_3/M^2)\Box\varphi \approx \frac{3}{4}\, k_2
(r_V/r)^{3/2} \gg k_2 > 0$. Similarly, the
$-(2k_3/M^2)\nabla^\mu\partial^\nu\varphi$ contribution to
$\mathcal{G}^{\mu\nu}$ gives at lowest post-Newtonian order
$\frac{1}{4}\, k_2 (r_V/r)^{3/2}$ for the $\mathcal{G}^{rr}$
component, minus twice this expression (multiplied by
$g^{\theta\theta}$ and $g^{\phi\phi}$) for the angular
$\mathcal{G}^{\theta\theta}$ and $\mathcal{G}^{\phi\phi}$
components, and again this expression but multiplied by a
negligible factor $r_S/r$ for the $\mathcal{G}^{00}$ component.
Finally, if we assume that $\alpha^2/k_2$ is at most of order one
($\alpha^2/k_2 \ll 1$ still being allowed), in order to avoid a
too large matter-scalar coupling in the asymptotic
$r\rightarrow\infty$ region, then the assumption $r_S \ll r \ll
r_V$ implies $(k_3\varphi'^2/M^2)^2 \approx (\alpha r_S/4 r)^2
\ll k_2 \ll (2k_3/M^2)\Box\varphi$, therefore the
$\mathcal{O}(k_3^2)$ terms entering the effective metric
(\ref{Eq16}) are fully negligible: They are of second
post-Newtonian order beyond the Vainshtein effect. In conclusion,
the effective metric (\ref{Eq16}) reads at lowest order
\begin{equation}
\mathcal{G}^{\mu\nu}_\text{Vainshtein} \approx
\frac{k_2}{4} \left(\frac{r_V}{r}\right)^{3/2}
\text{diag}\left(-3, 4, \frac{1}{r^2},
\frac{1}{r^2 \sin^2\theta}\right),
\label{Eq20}
\end{equation}
and is has therefore the right $-+++$ signature, warranting that
the Cauchy problem is well posed and that the scalar field
carries positive energy. The large numerical factor
$(r_V/r)^{3/2}$ implies that scalar perturbations are weakly
coupled to matter; this is a consequence of the Vainshtein
mechanism. The different integers entering the diagonal matrix
mean that the speed of scalar waves is $c/\sqrt{3}$ in the
orthoradial directions, but $2c/\sqrt{3}$ in the radial one. As
discussed in \cite{Bruneton:2006gf,Babichev:2007dw}, this
superluminal radial velocity does not violate causality precisely
because the Cauchy problem remains well posed simultaneously for
all the fields: At any spacetime point, there exists a
hypersurface which is spacelike with respect to both $g^{\mu\nu}$
and $\mathcal{G}^{\mu\nu}$, on which one may specify initial
data, and such hypersurfaces may be used to foliate the full
spacetime.

To conclude, the curved-spacetime Galileon model (\ref{Eq1}) in
an asymptotic Minkowskian Universe is consistent both in the
Vainshtein regime ($r\ll r_V$) and at large distances, provided
$k_2 >0$ (and $k_3\alpha > 0$ for the Vainshtein regime to
exist). For such an asymptotic Minkowski spacetime, the
background scalar field is given by solution (\ref{Eq8}),
(\ref{Eq9}) or (\ref{Eq12}) with their \textit{lower} sign.

This conclusion also assumes that $\alpha^2/k_2$ is not an
extremely large dimensionless number, i.e., that matter is not
too strongly coupled to the scalar field. If $|\alpha| \gg
(|k_3|/ M^2 r_S^2)^{1/3}$, then the $\mathcal{O}(\varphi'^4)$
contributions dominate in the effective metric (\ref{Eq16}) at
small enough distances, and its signature becomes $++--$ instead
of $-+++$. Scalar perturbations are thus ill behaved, not only
because their energy can be negative, but above all because their
field equation is not hyperbolic. However, the present study
assumed from the beginning that the backreaction of the scalar
field on the Schwarzschild metric was negligible, which is
obviously not the case in the very strong matter-scalar coupling
limit. Therefore, this non-hyperbolic signature does not even
prove the inconsistency of the Galileon model (\ref{Eq1}) in this
limit. It just means that we \textit{must} assume as above that
$\alpha^2/k_2$ is at most of order one, otherwise the scalar
backreaction on the metric cannot be neglected.

To estimate this backreaction, it suffices to compare the
scalar's energy-momentum tensor
\begin{eqnarray}
\frac{1}{M_P^2}\, T_{\mu\nu}(\varphi) &=&
k_2 \Bigl[2\,\partial_\mu\varphi\,\partial_\nu\varphi
- g_{\mu\nu}
\left(\partial_\lambda\varphi\right)^2\Bigr]
+\frac{k_3}{M^2}\Bigl[
2\,\Box\varphi\, \partial_\mu\varphi\,\partial_\nu\varphi
\nonumber\\
&&+ g_{\mu\nu}\partial_\rho\varphi\,
\nabla^\rho\left(\partial_\lambda\varphi\right)^2
- \partial_\mu\varphi\,
\nabla_\nu\left(\partial_\lambda\varphi\right)^2
- \partial_\nu\varphi\,
\nabla_\mu\left(\partial_\lambda\varphi\right)^2
\Bigr],
\label{Tmunu}
\end{eqnarray}
with the one of the matter source, or more specifically their
spatial integrals $\int_0^r \left(-T_0^0+T_i^i\right) 4\pi r'^2
dr'$, which generate the Newtonian potential $\frac{1}{2}(g_{00}
+1)$ at a distance $r$ from the center of the body. For any $r$,
one then finds that the scalar's contribution is
$\mathcal{O}\left( \alpha^{5/3} (M r_S)^{2/3}\right)$ smaller
than that of matter. Our analysis can thus be trusted only if
$\alpha$ is not too large, so that it does not compensate the
factor $(M r_S)^{2/3}$. On the other hand, even if the
matter-scalar coupling is of order one (i.e., $\alpha^2/k_2 \sim
1$), then the scalar's backreaction is negligible as soon as
$1/M$ is chosen large enough with respect to the Schwarzschild
radius of any body. In Galileon models, $M$ is generally assumed
to be of the order of the Hubble constant $H$ (and we will
actually derive so in Sec.~\ref{Sec6} below, but while assuming a
different asymptotic Universe). Then $M r_S$ is always an
extremely small number, about $10^{-23}$ for the Sun, $10^{-11}$
for a galaxy, and still $10^{-8}$ for a cluster. Therefore, our
test scalar field approximation is fully safe if $M\sim H$ and
$\alpha^2/k_2$ is at most of order one.

\section{Asymptotic de Sitter spacetime}
\label{Sec6}
Let us first consider an isotropic and homogeneous Universe,
described by the Friedmann-Lema\^{\i}tre-Robertson-Walker (FLRW)
metric $ds^2 = -d\tau^2 +
a(\tau)^2\left(d\rho^2+\rho^2d\Omega^2\right)$, where the
cosmological time $\tau$ and the comoving radius $\rho$ should
not be confused with the time $t$ and radial coordinate $r$ of
the Schwarzschild metric (\ref{Eq5}). Since action (\ref{Eq1})
does not involve any cosmological constant, the expansion of the
Universe will be caused by the Galileon field $\varphi$ itself.
As was shown in \cite{Deffayet:2010qz}, the combined Einstein and
scalar field equations indeed admit a self-accelerating solution.
Let us assume, like in Eq.~(\ref{Eq6}) above, that the scalar
field has a linear time dependence, $\varphi = \dot\varphi_c \tau
+\varphi_0$, i.e., that $\ddot\varphi = 0$ (where a dot denotes a
derivative with respect to $\tau$). Then the field equations read
\begin{subequations}
\label{Eq21}
\begin{eqnarray}
3 H^2 &=& \frac{\varepsilon}{M_P^2}
+ k_2 \dot\varphi_c^2 - 6 H\,\frac{k_3}{M^2}\,\dot\varphi_c^3,
\label{Eq21a}\\
-\dot H &=&\frac{\varepsilon+p}{2M_P^2}
+k_2\dot\varphi_c^2-3 H\,\frac{k_3}{M^2}\,\dot\varphi_c^3,
\label{Eq21b}\\
\frac{\partial_\tau\left(a^3 J^0\right)}{a^3}
&=& \alpha (\varepsilon-3p),
\label{Eq21c}
\end{eqnarray}
\end{subequations}
where $H\equiv \dot a/a$ is the Hubble parameter, and
$\varepsilon$ and $p$ denote the energy density and pressure
of matter, that we shall neglect in the following. [Actually,
Eqs.~(\ref{Eq21a}) and (\ref{Eq21c}) are valid even if
$\ddot\varphi\neq 0$.] The only
Christoffel contribution to the current (\ref{Eq3}) comes from
$\Box\varphi = -3H\dot\varphi_c$, and Eq.~(\ref{Eq21c}) reduces
to
\begin{equation}
\partial_\tau\left(k_2 a^3 \dot\varphi_c
-3 H \frac{k_3}{M^2}\,a^3 \dot\varphi_c^2\right) = 0,
\label{Eq22}
\end{equation}
which admits the solution $\dot\varphi_c = k_2M^2/3Hk_3$.
[In an expanding Universe, the other solutions, $J^0 =
\text{const}/a^3 \rightarrow 0$, tend either towards this
one or towards $\dot\varphi_c = 0$.] Then
Eq.~(\ref{Eq21b}) shows that the scalar-field analogue of
$\varepsilon+p$ vanishes, i.e., that it behaves as an effective
cosmological constant, and we consistently find that $H$ is also
constant. Finally, Eq.~(\ref{Eq21a}) reads $3 H^2 =
-k_2\dot\varphi_c^2$, implying that $k_2$ \textit{must} be
negative for this self-accelerating solution to exist, and it
gives the numerical value
\begin{equation}
H^2 = \frac{M^2}{|k_3|}\left(\frac{|k_2|}{3}\right)^{3/2}.
\label{Eq23}
\end{equation}
This allows us to rewrite $\dot\varphi_c$ in terms of the
constants entering action (\ref{Eq1}). Assuming an expanding
Universe (i.e., $H > 0$), we get
\begin{equation}
\dot\varphi_c = -\text{sign}(k_3)
\left(\frac{|k_2|}{3}\right)^{1/4} \frac{M}{\sqrt{|k_3|}}.
\label{Eq24}
\end{equation}

Let us now consider a static and spherically symmetric body
embedded in such an expanding Universe. As before, we assume that
the backreaction of the \textit{local} scalar field on the metric
is negligible, but its cosmological time evolution (\ref{Eq24})
is obviously taken into account since it is responsible for the
accelerated expansion. The local scalar-field solution can still
be written in the form (\ref{Eq8}), (\ref{Eq9}) or (\ref{Eq12}),
at least at small enough distances, and the effective metric in
which scalar perturbations propagate is still given by
Eq.~(\ref{Eq16}). We saw in Sec.~\ref{Sec5} that the scalar
perturbations are ghostlike at large distances if $k_2 < 0$, but
this was derived while assuming asymptotic flatness and thereby
the Brans-Dicke like behavior (\ref{Eq19}) of the background
scalar field at infinity. Therefore this previous result is no
longer valid in the present de Sitter Universe. However, the
small-distance physics remains \textit{a priori} unchanged,
notably within the Vainshtein radius, and we saw that the
signature of the effective metric (\ref{Eq16}) depended on the
sign of $k_3 \Box\varphi$. As mentioned below Eq.~(\ref{Eq12}),
this sign actually does not depend on $k_3$, but it does depend
on $k_2$, and Eq.~(\ref{Eq20}) confirms that we \textit{a priori}
get a ghostlike scalar perturbation if $k_2 < 0$. The stability
of the self-accelerating solution seems thus spoiled by the
presence of any massive body.

A first way out would be to consider the fine-tuned model such
that $\alpha_\text{eff} = 0$. Then the scalar field would not be
perturbed at all by local matter, there would not exist any
Vainshtein regime, and the proof in Ref.~\cite{Deffayet:2010qz}
that the model is stable would then be valid. This assumption
$\alpha_\text{eff} = 0$ actually seems more natural in the
present self-accelerating Universe, since the condition
$\dot\varphi_c = (-\alpha/k_3)^{1/2}M$ derived at the end of
Sec.~\ref{Sec3} translates as $\alpha =
-\text{sign}(k_3)\sqrt{|k_2|/3}$, involving only constant
parameters. However, if we take into account the matter sources
in Eqs.~(\ref{Eq21}), to derive a more realistic expansion of the
Universe, then neither $H$ nor $\dot\varphi_c$ will remain
constant, and we will eventually reach an epoch for which
$\alpha$ is no longer tuned to the right numerical value.
Moreover, the self-gravity effects mentioned in footnote
\ref{foot2} mean that the bare coupling constant $\alpha$ needs
to be replaced by a body-dependent product $\alpha (1-2s)$, which
cannot be fine-tuned to $-\text{sign}(k_3)\sqrt{|k_2|/3}$ for all
bodies. Therefore, even for this specific model, one expects that
the scalar field will be directly coupled to matter with a
nonvanishing $\alpha_\text{eff}$, at least at some cosmological
epoch, and this seems to generically lead to ghost instabilities
because $k_2 < 0$.

In fact, the small-distance effective metric (\ref{Eq20}) is
\textit{not} correct in the present expanding Universe. Indeed,
solution (\ref{Eq12}) does depend on the sign of $k_2$, but also
on the global $\mp$ sign. We did prove in Sec.~\ref{Sec5} that
the lower ($+$) sign gave the correct solution in an
asymptotically Minkowskian Universe, but this is no longer the
case in a de Sitter one. This comes from the fact that the FLRW
coordinates $\tau$ and $\rho$, useful at cosmologically large
distances, do not coincide with the $t$ and $r$ coordinates
defining the static and spherically-symmetric local metric
(\ref{Eq5}). In presence of a cosmological constant $\Lambda = 3
H^2$, here mimicked by the self-accelerating solution
(\ref{Eq23}), we do know that the exact Schwarzschild-de Sitter
solution can be written as Eq.~(\ref{Eq5}) with
\begin{equation}
e^\nu = e^{-\lambda} = 1-r_S/r - (H r)^2.
\label{Eq25}
\end{equation}
This can be matched to an asymptotic FLRW coordinate system by
defining
\begin{equation}
\begin{aligned}[c]
\tau&=t+\frac{1}{2H}\ln\left[1-(Hr)^2\right],\\
\rho&=\frac{e^{-Ht}}{\sqrt{1-(Hr)^2}}\, r\,,
\end{aligned}
\qquad\Leftrightarrow\qquad
\begin{aligned}[c]
t&=\tau-\frac{1}{2H}\ln\left[1
-\left(H e^{H\tau}\rho\right)^2\right],\\
r&=e^{H\tau}\rho\,.
\end{aligned}
\label{Eq26}
\end{equation}
Let us also set
\begin{equation}
B\equiv 1 - \frac{r_S}{e^{H\tau} \rho}
- \left(H e^{H\tau} \rho \right)^2.
\label{Eq27}
\end{equation}
Then the Schwarzschild-de Sitter metric (\ref{Eq5})-(\ref{Eq25})
becomes
\begin{subequations}
\label{Eq28}
\begin{eqnarray}
ds^2 &=& -B
\left[\frac{d\tau+H e^{2H\tau}\rho\,
d\rho}{1-\left(H e^{H\tau}\rho\right)^2}\right]^2
+ e^{2H\tau}\left[\frac{\left(d\rho+H\rho\,
d\tau\right)^2}{B} + \rho^2 d\Omega^2\right]
\label{Eq28a}\\
&=&-\left[1-\frac{r_S}{H^2\left(e^{H\tau}
\rho\right)^3}+\mathcal{O}
\left(\frac{1}{\rho^4}\right)\right]d\tau^2
+\left[\frac{4r_S}{H^3e^{3 H\tau}\rho^4}
+\mathcal{O}\left(\frac{1}{\rho^5}\right)\right]d\tau\, d\rho
\nonumber\\
&&+e^{2H\tau}
\left\{\left[1+\frac{r_S}{H^2\left(e^{H\tau}\rho\right)^3}
+\mathcal{O}\left(\frac{1}{\rho^4}\right)\right]d\rho^2
+\rho^2 d\Omega^2\right\},
\label{Eq28b}
\end{eqnarray}
\end{subequations}
in which one recognizes the asymptotic de Sitter metric $ds^2 =
-d\tau^2 + e^{2H\tau}\left(d\rho^2+\rho^2d\Omega^2\right)$ up to
corrections of order $\mathcal{O}(r_S)$ caused by the local
massive body. Note that we merely changed coordinates, and that
Eq.~(\ref{Eq28}) still defines the same \textit{static} solution
as (\ref{Eq5})-(\ref{Eq25}), in spite of the presence of
time-dependent exponentials. In particular, the Schwarzschild
radius still corresponds to the constant $e^{H\tau}\rho = r_S$,
the factor $e^{H\tau}$ compensating the fact that we now measure
lengths with a varying (comoving) ruler. Note also that the
spherical body is located at $\rho = 0$ at any time, therefore it
is comoving (actually static) in the background de~Sitter
Universe.

To derive the cosmological solution (\ref{Eq23})-(\ref{Eq24}), we
assumed that the scalar field is homogeneous at large distances
in de~Sitter coordinates, i.e., that it reads $\varphi =
\dot\varphi_c \tau +\varphi_0$ without any comoving radius
($\rho$) dependence. This means that in terms of the
Schwarzschild coordinates $t$ and $r$, Eqs.~(\ref{Eq26}),
$\varphi$ does depend on $r$. We explicitly get
\begin{eqnarray}
\varphi &=& \dot\varphi_c t
+\frac{\dot\varphi_c}{2H}\ln\left[1-(Hr)^2\right]
+ \varphi_0
\nonumber\\
\Rightarrow\quad
\varphi'&=& -\dot\varphi_c\, \frac{Hr}{1-(Hr)^2}
= -\frac{k_2 M^2}{3 k_3}\,\frac{r}{1-(Hr)^2},
\label{Eq29}
\end{eqnarray}
$H^2$ being given by Eq.~(\ref{Eq23}). In other words, the
correct solution (\ref{Eq8}) for the background scalar field
$\varphi'$, embedded in a self-accelerating Universe, should
contain a local $r$ dependence, and with the precise factor
$-k_2M^2/3k_3$ entering Eq.~(\ref{Eq29}). Let us prove that this
result is given by the \textit{upper} sign of (\ref{Eq8}).
Indeed, this solution can no longer be expanded as (\ref{Eq9})
for too large radii $r$, because expressions (\ref{Eq25}) for
$\nu(r)$ and $\lambda(r)$ must now be used, so that $\nu' e^{\nu}
= -\lambda' e^{\nu} = r_S/r^2 - 2 H^2 r$. We get for such large
distances (still assumed smaller than $1/\sqrt{3}H$)
\begin{equation}
\varphi' =
-\frac{k_2 M^2}{4 k_3}\,
\frac{r}{1-\frac{3}{2}(Hr)^2}
\left[1\pm\frac{1}{3}\,\frac{1-3(Hr)^2}{1-(Hr)^2}
\right]\times
\left[1+\mathcal{O}\left(\frac{r_S}{r}\right)\right],
\label{Eq30}
\end{equation}
therefore the upper sign recovers \textit{exactly} the asymptotic
behavior derived in Eq.~(\ref{Eq29}) from a matching with the
cosmological solution, whereas the lower sign gives this result
divided by $\left[2-3(Hr)^2\right]$, i.e., with not only an
erroneous factor $\frac{1}{2}$ by also an incorrect radial
dependence.

Solution (\ref{Eq8}), with its correct upper sign, may now be
written for $Hr\ll1$ as
\begin{equation}
\varphi' =
-\frac{k_2 M^2r}{4 k_3}
\left(1+\sqrt{1-\frac{8}{9}+\frac{4 k_3 r_S}{k_2^2 M^2 r^3}\,
\alpha_\text{eff}}\right)\times
\left[1+\mathcal{O}\left(\frac{r_S}{r}\right)
+\mathcal{O}\left(H^2r^2\right)\right],
\label{Eq31}
\end{equation}
where the constant $-\frac{8}{9}$ within the square root comes
{}from a compensation between the $r_S/r^3$ coefficient entering
(\ref{Eq8}) and the large-distance behavior of $\nu' r^2/r_S$,
while using the values (\ref{Eq23}) and (\ref{Eq24}) for $H^2$
and $\dot\varphi_c$. Note that this solution is different from
Eq.~(\ref{Eq9}), which was valid for an asymptotically
Minkowskian spacetime. In particular, (\ref{Eq31}) does not
reduce to the Brans-Dicke solution (\ref{Eq19a}) for $r \gg r_V$
(while still assuming $r\ll 1/H$), but gives
\begin{equation}
\varphi' =
-\left(\frac{k_2 M^2r}{3 k_3}
+\frac{3\,\alpha_\text{eff}\, r_S}{2k_2r^2}\right)\times
\left[1+\mathcal{O}\left(\frac{r_S}{r}\right)
+\mathcal{O}\left(\frac{r_V^6}{r^6}\right)
+\mathcal{O}\left(H^2r^2\right)\right],
\label{Eq32}
\end{equation}
with different sign and numerical coefficient for the usual
$\mathcal{O}(1/r^2)$ contribution, but above all involving a much
larger $\mathcal{O}(r)$ term, that we already found in
Eq.~(\ref{Eq29}) [it is $\mathcal{O}\left(r^3/r_V^3\right)$
larger than the $\mathcal{O}(1/r^2)$ contribution].

In the Vainshtein regime ($r_S \ll r \ll r_V$), on the other
hand, we still get Eq.~(\ref{Eq12}), but with the crucial
information that the upper sign must be chosen:
\begin{equation}
\varphi' =
-\frac{k_2 M^2}{4 k_3}
\sqrt{\frac{r_V^3}{r}}
\left[1+\mathcal{O}\left(\frac{r_S}{r}\right)
+\mathcal{O}\left(\frac{r^3}{r_V^3}\right)
+\mathcal{O}\left(H^2r^2\right)\right].
\label{Eq33}
\end{equation}
This is the opposite expression as the one we used in
Sec.~\ref{Sec5} for an asymptotically Minkowskian spacetime, and
the effective metric in which scalar perturbations propagate, at
small distances, is thus the opposite of (\ref{Eq20}) [up to
negligible corrections of relative order
$\mathcal{O}(\sqrt{r_S/r})$ or $\mathcal{O}(\sqrt{r^3/r_V^3})$,
due to the nonvanishing time derivative $\dot\varphi_c$ we now
must take into account when computing (\ref{Eq16})]. Since we
know that $k_2 < 0$ for the present self-accelerating Universe to
be a solution, this effective metric is therefore again of the
correct signature $-+++$, warranting a well-posed Cauchy problem
and the absence of ghosts.

The signature of this effective metric must also be studied at
large distances $r \gg r_V$. If one neglects all contributions
proportional to the Schwarzschild radius $r_S$ of the massive
body, $\mathcal{G}^{\mu\nu}$ is most conveniently computed in
FLRW coordinates. Then one finds that all contributions to
Eq.~(\ref{Eq16}) are of the same order of magnitude
$\mathcal{O}(k_2)$, including those of the usually negligible
$(\partial\varphi)^4$ terms. The effective metric finally reads
$\mathcal{G}^{\mu\nu} = \text{diag}(-2|k_2|,0,0,0) +
\mathcal{O}(r_S)$, meaning that the Cauchy problem remains well
posed for scalar perturbations, but that their sound velocity
vanishes, as was already noticed in \cite{Deffayet:2010qz}. To
avoid any instability, the $\mathcal{O}(r_S)$ corrections should
thus contribute positively to the effective spatial metric
$\mathcal{G}^{ij}$. We thus need to compute them, while taking
into account the zeroth-order terms $\mathcal{O}(r_S^0)$ although
we know they will eventually cancel. Indeed some of these
zeroth-order terms get multiplied by $r_S$ when solution
(\ref{Eq8}) is expanded to compute (\ref{Eq16}). Another
complication is that $\mathcal{G}^{0r} \neq 0$ in Schwarzschild
coordinates (\ref{Eq5})-(\ref{Eq25}), i.e., that
$\mathcal{G}^{\mu\nu}$ is not diagonal. However, to determine its
signature, it suffices to consider the quadratic form
$\mathcal{G}_{\mu\nu} dx^\mu dx^\nu$ (where the effective
metric's indices are lowered with $g_{\mu\nu}$, i.e.,
$\mathcal{G}_{\mu\nu}$ is \textit{not} the inverse of
$\mathcal{G}^{\mu\nu}$), and to note that
$\mathcal{G}_{00} dt^2 + 2 \mathcal{G}_{0r} dt dr
+ \mathcal{G}_{rr} dr^2 = \mathcal{G}_{00}
\left(dt+ \mathcal{G}_{0r} dr/\mathcal{G}_{00}\right)^2
+ \left(\mathcal{G}_{rr}
- \mathcal{G}_{0r}^2/\mathcal{G}_{00}\right) dr^2$.
In this new frame diagonalizing the effective metric, one finally
finds at lowest order $\mathcal{G}_{\mu\nu} \approx
\text{diag}\left( -2|k_2| , 2 |k_2| \epsilon, -|k_2| \epsilon
r^2, -|k_2| \epsilon r^2\sin^2\theta \right)$, where $\epsilon
\equiv \frac{3}{4}(r_V/r)^3$, i.e., a signature $-+--$ instead of
$-+++$, implying that the Cauchy problem is ill-posed for scalar
perturbations propagating in the orthoradial directions (or that
they are unstable if one interprets these negative signs as an
imaginary sound speed, such instabilities being \textit{large} at
moderately large distances of a few $r_V$). In conclusion,
although we found above that the physics of scalar perturbations
is consistent in the Vainshtein regime $r \ll r_V$, it is
\textit{not} at large distances $r \gg r_V$.

This serious problem can \textit{a priori} have several
solutions. The first naive idea would be to choose the theory
parameters such that $r_V \agt 1/H$, so that any instability
would be hidden behind the cosmological horizon. But since $r_V
\propto r_S^{1/3}$, this cannot work for all massive bodies, and
this would need anyway a very large matter-scalar coupling
constant $\alpha$, thereby spoiling the hyperbolicity in the
Vainshtein regime for the same reasons as at the end of
Sec.~\ref{Sec5} (and being inconsistent with our assumption of a
negligible backreaction of the local scalar field on the
background metric). A much better argument is that the scalar
field may never get out of the Vainshtein regime if one takes
into account the full matter distribution in the Universe: At any
location, there would be a close enough and massive enough body
such that $r < r_V$. But the strongest argument is that the
vanishing sound speed found above at order $\mathcal{O}(r_S^0)$
is a consequence of the exact self-accelerating solution
(\ref{Eq23})-(\ref{Eq24}). As soon as one takes into account
matter in the field equations (\ref{Eq21}), then the sound speed
becomes nonvanishing and real, as was shown in
\cite{Deffayet:2010qz}, and it dominates at large distances over
any $\mathcal{O}(r_S)$ correction. For instance, if we assume
$\alpha = 0$ like in \cite{Deffayet:2010qz}, the presence of a
positive matter density $\varepsilon$ in Eq.~(\ref{Eq21a})
increases the value of $H$ with respect to the lowest-order
expression (\ref{Eq23}). This implies a modification of
Eq.~(\ref{Eq24}) for $\dot\varphi_c$, and the matter field
equation $\dot\varepsilon + 3 H \varepsilon = 0$ (assuming a
vanishing pressure $p$) allows us to compute $\ddot\varphi_c$.
We finally find that the terms proportional to $\Box\varphi$ and
$\left(\partial_\lambda\varphi\right)^4$ in Eq.~(\ref{Eq16}) both
give \textit{positive} $\mathcal{O}(\varepsilon)$ contributions
to the spatial effective metric $\mathcal{G}^{ij}$, and we
explicitly obtain
\begin{equation}
\mathcal{G}^{ij} = \frac{5}{18}\, |k_2|\, g^{ij}
\frac{\varepsilon}{M_P^2 H^2}
+\mathcal{O}\left(\varepsilon^2\right).
\label{GijNoAlpha}
\end{equation}
Surprizingly, the result is fully different if we assume a
nonzero bare coupling constant $\alpha$, because the matter field
equation $\dot\varepsilon + 3 H \varepsilon = \alpha \varepsilon
\dot\varphi_c$ implies a different time evolution for
$\varepsilon$, and the time integration of Eq.~({\ref{Eq21c})
makes $\alpha$ go away. One finds that the lowest-order
expressions of $H$ and $\dot\varphi_c$ both get multiplied by the
same factor $1 - \varepsilon/12 M_P^2 H^2 +
\mathcal{O}(\varepsilon^2)$, and the matter field equation can
again be used to compute $\ddot\varphi_c$. Now the term
proportional to $\Box\varphi$ in Eq.~(\ref{Eq16}) happens to give
a negative $\mathcal{O}(\varepsilon)$ contribution to
$\mathcal{G}^{ij}$, but it is counterbalanced by the
$\left(\partial_\lambda\varphi\right)^4$ and
$\nabla^\mu\partial^\nu\varphi$ terms of (\ref{Eq16}).
We finally get
\begin{equation}
\mathcal{G}^{ij} = \frac{1}{18}\, |k_2|\, g^{ij}
\frac{\varepsilon}{M_P^2 H^2}
\left(1+ \text{sign}(k_3) \alpha \sqrt{\frac{3}{|k_2|}}\right)
+\mathcal{O}\left(\varepsilon^2\right),
\label{Gij}
\end{equation}
which is positive definite if $3\alpha^2/|k_2| < 1$ (and for even
larger $|\alpha|$ if $k_3 \alpha > 0$). The reason why
(\ref{Gij}) does not tend to (\ref{GijNoAlpha}) when $\alpha
\rightarrow 0$ is because they correspond to different
cosmological initial conditions. In any case, their positivity
shows that the ill-posed Cauchy problem found above was the
consequence of an oversimplified cosmological background, and
the signature of the effective metric at large distances is in fact
$-+++$, as needed.

To conclude, the curved-spacetime Galileon model (\ref{Eq1}) with
$k_2 < 0$ admits a self-accelerating cosmological solution, and a
spherical body embedded in this Universe generates a scalar field
given by the \textit{upper} sign of Eq.~(\ref{Eq8}), i.e., by
(\ref{Eq31}), (\ref{Eq32}) or (\ref{Eq33}) depending on the
distance to this body ---~the Vainshtein regime (\ref{Eq33})
existing only if $k_3 \alpha_\text{eff} > 0$. Perturbations
around this solution carry positive energy and have a well-posed
Cauchy problem in the Vainshtein regime $r \ll r_V$. Farther
away, one must take into account the matter content of the
Universe to derive the cosmological expansion, and the effective
metric in which scalar perturbations propagate then remains of
the correct hyperbolic signature $-+++$.

Let us finally mention in which conditions the local scalar
field's backreaction on the metric is indeed negligible, as was
assumed to draw the above conclusions. First, this is always the
case if $\alpha_\text{eff}$ is small enough, i.e., if one tunes
the bare matter-scalar coupling constant to $\alpha \approx
-\text{sign}(k_3)\sqrt{|k_2|/3}$, so that it almost compensates
the cosmologically induced coupling $k_3\dot\varphi_c^2/M^2$
entering Eq.~(\ref{Eq10}). Indeed, the spatial integral of
Eq.~(\ref{Tmunu}) over a sphere of radius $r < 1/H$ is at most of
order $\mathcal{O}\left(\alpha_\text{eff}\, r_S\right)$,
negligible with respect to the effect of the material body of
mass $\frac{1}{2}\,r_S$. However, for a random bare coupling
constant such that $\alpha^2/|k_2| \alt 1$ (a small value being
also allowed), Eqs.~(\ref{Eq10}) and (\ref{Eq24}) imply that
$\alpha_\text{eff}^2/|k_2|$ is generically of the order of unity,
therefore one actually expects large deviations from our results
above. But even in such a case, the scalar field's backreaction
is still negligible in the Vainshtein regime $r \ll r_V$, because
it is of order $\mathcal{O}\left(\alpha_\text{eff}\, r_S
(r/r_V)^{3/2}\right)$. Therefore, our previous conclusion that
the Galileon model (\ref{Eq1}) is stable in the Vainshtein regime
remains valid even for a matter-scalar coupling of order one. On
the other hand, for such a $\mathcal{O}(1)$ coupling, our test
scalar field approximation cannot be trusted at distances $r \agt
r_V$, and further work is needed to actually prove
stability~\cite{BabichevEtAl}. Finally, for cosmologically large
distances $r \sim 1/H \gg r_V$, both the effects of the massive
body and the locally generated scalar field are negligible, and
we asymptotically reach an effective metric $\mathcal{G}^{\mu\nu}
\approx |k_2| \text{diag}\left(-2, \mathcal{O}(1)\, \varepsilon\,
g^{ij}/M_P^2 H^2 \right)$, with the correct hyperbolic signature
$-+++$.

\section{Scalar waves within matter}
\label{Sec7}
Although we have shown in Secs.~\ref{Sec5} and \ref{Sec6} that
the field equations for scalar perturbations are hyperbolic in
the Vainshtein regime, and thereby that there is no ghost
instability at small enough distances from a massive body, the
growth of such perturbations \textit{within} the body deserves a
more careful study. First of all, the vacuum solution (\ref{Eq8})
for the scalar field is no longer valid inside matter, therefore
the hyperbolicity of the effective metric (\ref{Eq16}) needs to
be checked. To simplify, let us consider a spherical body of
constant density. As derived in page 331 of Ref.~\cite{Weinberg},
the interior solution for the metric (\ref{Eq5}) (neglecting
cosmological corrections of order $H^2 r^2$) reads
\begin{subequations}
\label{Eq34}
\begin{eqnarray}
e^{\nu(r)} &=&
\frac{1}{4}\left[3\left(1-\frac{r_S}{r_*}\right)^{1/2}
-\left(1-\frac{r_S r^2}{r_*^3}\right)^{1/2}\right]^2
= 1 - \left(3 - \frac{r^2}{r_*^2}\right) \frac{r_S}{2r_*}
+\mathcal{O}\left(r_S^2\right),
\label{Eq34a}\\
e^{\lambda(r)} &=& \left(1-\frac{r_S r^2}{r_*^3}\right)^{-1}
= 1 + \frac{r_S r^2}{r_*^3}
+\mathcal{O}\left(r_S^2\right),
\label{Eq34b}
\end{eqnarray}
\end{subequations}
where $r_*$ denotes the body's radius, not to be confused with
its Schwarzschild radius $r_S$. The background scalar field
equation (\ref{Eq2}) can then be solved as in Sec.~\ref{Sec3},
with the difference that only the mass interior to the sphere of
radius $r$, $m(r) = (r_S/2G) (r/r_*)^3$, is a source for
$\varphi'(r)$. We get
\begin{equation}
\varphi'_\text{interior} = -\frac{k_2 M^2 r}{4 k_3}
\left(1\pm\sqrt{1-\frac{8 H^2 k_3^2}{k_2^2 M^4}\,
\dot\varphi_c^2
+\frac{4 k_3 r_S}{k_2^2
M^2 r_*^3}\, \alpha_\text{eff}}\right)
+ \mathcal{O}\left(r_S^2\right)
+ \mathcal{O}\left(H^2 r^2\right),
\label{Eq35}
\end{equation}
where either $H = 0$ in asymptotic Minkowski spacetime, or $H$
and $\dot\varphi_c$ are given by Eqs.~(\ref{Eq23}) and
(\ref{Eq24}) in asymptotic de Sitter spacetime (yielding the
constant $-\frac{8}{9}$ we already encountered in
Eq.~(\ref{Eq31}) for the exterior solution). The only difference
with respect to the exterior solution (\ref{Eq9}) or
(\ref{Eq31}), at this lowest post-Newtonian order, is that
$\alpha_\text{eff}$ is multiplied by the constant $1/r_*^3$
instead of $1/r^3$. In the Vainshtein regime $r\leq r_* \ll r_V$,
we have thus
\begin{equation}
\varphi'_\text{interior} = \mp\frac{k_2 M^2}{4 k_3}
\left(\frac{r_V}{r_*}\right)^{3/2} r
\left[1+\mathcal{O}\left(\frac{r_S}{r_*}\right)
+\mathcal{O}\left(\frac{r_*^3}{r_V^3}\right)
+\mathcal{O}\left(H^2r_*^2\right)\right],
\label{Eq36}
\end{equation}
instead of Eq.~(\ref{Eq12}), where the lower sign ($+$)
corresponds to asymptotic Minkowski spacetime, and the upper one
($-$) to asymptotic de Sitter spacetime. The various
contributions to the effective metric (\ref{Eq16}) can now be
computed within the body, and one finds at lowest order (in both
asymptotic Minkowski and de Sitter cases)
\begin{equation}
\mathcal{G}^{\mu\nu}_\text{interior} \approx
|k_2| \left(\frac{r_V}{r_*}\right)^{3/2}
\text{diag}\left(-\frac{3}{2}, 1, \frac{1}{r^2},
\frac{1}{r^2 \sin^2\theta}\right),
\label{Eq37}
\end{equation}
instead of (\ref{Eq20}) [or the opposite of of (\ref{Eq20}) in
the asymptotic de Sitter case, for which $-k_2 = |k_2|$]. Note
that this effective metric is \textit{discontinuous} at the
surface of the body, because it involves $\varphi''$, which is
discontinuous when one suddenly passes from empty space to a
finite matter density. But of course, everything becomes
continuous when one considers a smooth matter profile. Note also
that the sound speed, $\sqrt{2/3}\, c$, is now isotropic, and
differs from its value outside matter (that we found in
Sec.~\ref{Sec5} to be $2c/\sqrt{3}$ in the radial direction and
$c/\sqrt{3}$ in the orthoradial ones). This can be interpreted as
different refractive indices of matter and vacuum for scalar
waves. But the crucial information brought by Eq.~(\ref{Eq37}) is
that the effective metric remains of hyperbolic signature $-+++$
within matter, therefore no ghost instability develops even in
the interior of a body.

On the other hand, the Ricci tensor entering Eq.~(\ref{Eq4}) no
longer vanishes within matter, and this causes extra couplings of
the perturbations to the background fields, whose effects could
\textit{a priori} lead to other types of instabilities. This
Ricci tensor is actually generated both by matter and by the
background scalar field itself. The scalar-scalar
self-interactions are responsible for the $\mathcal{O}(k_3^2)$
terms in the effective metric (\ref{Eq16}), as was first shown in
\cite{Deffayet:2010qz}. Not only we already took them into
account to derive (\ref{Eq37}), but we also saw in
Secs.~\ref{Sec5} and \ref{Sec6} that they are negligible with
respect to the $\mathcal{O}(k_3)$ terms, provided
$\alpha^2/|k_2|$ is at most of order 1 (i.e., that we are not
considering a extremely large matter-scalar coupling constant).
We can thus focus on the Ricci tensor generated by matter,
$R^{\mu\nu}_\text{matter} = \frac{3}{2}\left(r_S/r_*^3\right)
g^{\mu\nu} + \mathcal{O}\left(r_S^2\right)$, which multiplies
\textit{first} derivatives of the scalar field in
Eq.~(\ref{Eq4}), and thereby does not change its kinetic term
(second derivatives). Nevertheless, its presence means that
within matter, scalar perturbations acquire a direct derivative
coupling to the background scalar field. The first-order
expansion of (\ref{Eq4}) reads
\begin{equation}
\Box\varphi\, \Box\pi
- \left(\nabla^\mu\partial^\nu\varphi\right)
\left(\nabla_\mu\partial_\nu\pi\right)
-R^{\mu\nu}_\text{matter}\, \varphi_{,\mu}\pi_{,\nu} = 0,
\label{Eq38}
\end{equation}
where we have neglected the $\mathcal{O}(k_3^2)$ terms coming
{}from the diagonalization (\ref{Eq14})--(\ref{Eq16}) of the
kinetic terms, as well as the $\left(k_2M^2/2k_3\right) \Box\pi$
linear term because we assume the star is in the Vainshtein
regime ($r_* \ll r_V$). At lowest post-Newtonian order, the
radial contribution $-R^{rr}_\text{matter}\, \varphi' \pi'$ can
also be neglected, because both $R^{rr}_\text{matter} =
\mathcal{O}\left(r_S\right)$ and $\varphi' =
\mathcal{O}\left(\sqrt{r_S}\right)$ tend to $0$ with $r_S$. But
the temporal contribution $-R^{00}_\text{matter}\,
\dot\varphi_c\, \dot\pi$ is crucial, as it corresponds to a
damping or antidamping term depending on its sign. Let us thus
consider the asymptotic de Sitter case of Sec.~ \ref{Sec6}, where
$\dot\varphi_c$ is given by Eq.~(\ref{Eq24}). The main
contributions to Eq.~(\ref{Eq38}) then read
\begin{eqnarray}
\nabla_\mu\left(\mathcal{G}^{\mu\nu}_\text{interior}
\partial_\nu\pi\right)
-\frac{2 k_3}{M^2}\,R^{00}_\text{matter}\,
\dot\varphi_c\, \dot\pi &=& 0
\nonumber\\
\Leftrightarrow \quad
-\ddot\pi+\frac{2}{3}\,\Delta\pi
-\left(\frac{|k_2|}{3\, \alpha_\text{eff}^2}\right)^{1/4}
\left(\frac{r_S}{r_*^3}\right)^{1/2}
\dot\pi &=& 0.
\label{Eq39}
\end{eqnarray}
The coefficient of the $\dot\pi$ term is generically of order
$\sqrt{r_S/r_*^3}$, notably when the bare matter-scalar coupling
constant $\alpha$ vanishes or is negligibly small. Indeed,
Eqs.~(\ref{Eq10}) and (\ref{Eq24}) then give $\alpha_\text{eff}^2
\approx |k_2|/3$. Note that this coefficient can never be small
with respect to $\sqrt{r_S/r_*^3}$, because we assume that
$\alpha^2/|k_2|$ is at most of order 1, therefore
$\alpha_\text{eff}^2/|k_2|$ is never large either. On the other
hand, it is possible to choose the bare $\alpha$ so that
$\alpha_\text{eff}$ is small (see Sec.~\ref{Sec6}), and the
$\dot\pi$ term of Eq.~(\ref{Eq39}) may thus be multiplied by a
large number in some specific situations. Therefore, the extra
coupling of the perturbation $\pi$ to the cosmologically-imposed
$\dot\varphi_c$ \textit{within} matter can change drastically its
behavior. The good news is that its sign implies this is a
friction term, and therefore that no scalar instability occurs
within the material body. The plane-wave solutions of
Eq.~(\ref{Eq39}) read indeed (in Cartesian coordinates)
\begin{equation}
\pi \propto e^{-t/T}\sin\left(\omega t
- \mathbf{k}\cdot\mathbf{x}+\text{const}\right),
\label{Eq40}
\end{equation}
with a dispersion relation
\begin{equation}
\frac{2}{3}\, \mathbf{k}^2 = \omega^2 + \frac{1}{T^2},
\label{Eq41}
\end{equation}
and a decay time
\begin{equation}
T \equiv
2\left(\frac{3\, \alpha_\text{eff}^2}{|k_2|}\right)^{1/4}
\left(\frac{r_*^3}{r_S}\right)^{1/2}.
\label{Eq42}
\end{equation}
For the Sun, $2\sqrt{r_*^3/r_S}$ is about one hour, therefore
this damping is always quite efficient. But when
$\alpha_\text{eff}$ happens to be small, because of a balance
between the bare matter-scalar coupling constant $\alpha$ the
cosmologically-induced one $k_3\dot\varphi_c^2/M^2$ in
Eq.~(\ref{Eq10}), scalar perturbations are suppressed even
faster. For $\omega = 0$ and $\frac{2}{3}\,\mathbf{k}^2 \leq
1/T^2$, there exist other solutions of the form $\pi \propto
e^{-t/T_\pm}\sin\left(\mathbf{k}\cdot\mathbf{x}
+\text{const}\right)$, where $1/T_\pm \equiv 1/T \pm \sqrt{1/T^2
- 2\mathbf{k}^2/3}$ is always positive, therefore such
perturbations are damped too. The upper sign gives an even faster
decay than for the plane waves (\ref{Eq40}), whereas the lower
sign corresponds to a slower damping. The limiting case of a
homogeneous perturbation ($\mathbf{k} = \mathbf{0}$) gives either
a fast damping $\pi \propto e^{-2t/T}$, or an actual
constant which can be reabsorbed in the definition of the
background $\varphi_0$.

In conclusion, the physics of scalar perturbations is fully safe
in the interior of a spherical body in the Vainshtein regime. Not
only the Cauchy problem is well posed and the kinetic energy
carried by scalar perturbations is positive, but even the subtle
derivative coupling caused by the nonvanishing Ricci tensor in
Eq.~(\ref{Eq38}) efficiently \textit{damps} these perturbations.
Note that this local friction within material bodies is
paradoxically caused by the slow cosmological time evolution
($\dot\varphi_c \neq 0$) of the background scalar field. The
large-distance behavior of the Universe is thus significantly
influencing local physics.

\section{Conclusions}
\label{Concl}

In this paper we studied in detail spherically symmetric
solutions for the cubic covariant Galileon model (\ref{Eq1}) in
presence of a matter source. An important ingredient of our study
is that we take into account the time variation of the scalar
field, induced by the cosmological evolution. Although
cosmological time evolution is tiny (normally of order of the
Hubble rate, $\dot\varphi_c \sim H$), nevertheless its effect on
the behavior of the scalar is crucial and cannot be disregarded.

We exhibited, in particular, an {\it induced} matter-scalar
coupling for $\dot\varphi_c \neq0$, Eq.~(\ref{Eq10}). This effect
arises because of the kinetic mixing of the Galileon and the
graviton, which results in the coupling of the derivatives of the
scalar field to the curvature, in the Galileon field equation.
For $\dot\varphi_c \neq0$, this term plays the role of a
matter-scalar coupling, hence it is {\it induced} by the
cosmological evolution of $\varphi$. It is important to note that
this coupling is naturally of order of unity, therefore the
Galileon effectively couples to matter sources even if the bare
scalar-matter coupling is absent, $\alpha =0$.

We also found that the local solution for the Galileon field
crucially depends on the asymptotic conditions at large
distances. Indeed, to find time-dependent solutions in presence
of massive sources, we applied the ansatz (\ref{Eq6}), which
enables to separate the time and radial variables. The
radial-dependent part can then be found exactly, in the test
scalar field approximation, for an arbitrary static background
metric, Eq.~(\ref{Eq8}). [We checked \textit{a posteriori} the
conditions for backreaction to be negligible.] The full solution
for $\varphi'$, however, consists of two different branches. The
solution with the {\it lower} sign in (\ref{Eq8}) corresponds to
the asymptotically Minkowski spacetime, with no time evolution of
$\varphi$. The other case we studied in detail is the
asymptotically de Sitter Universe, whose self-acceleration is
provided by a time evolution of the Galileon itself. In this
case, the cosmological asymptotic behavior dictates to choose the
other branch of the solution, namely, Eq.~(\ref{Eq8}) with the
{\it upper} sign. It turns out that the stability of the
solutions drastically depends of the choice of the branch.

In order to investigate stability of the found solutions, notably
the well-posedness of the Cauchy problem and the positivity of
energy, we identified the actual spin-0 and spin-2 degrees of
freedom by diagonalizing their kinetic terms in Sec.~\ref{Sec4}.
Using the results of Sec.~\ref{Sec4}, we found in Sec.~\ref{Sec5}
that the background scalar field given by solution (\ref{Eq8}),
(\ref{Eq9}) or (\ref{Eq12}) with their \textit{lower} sign
(corresponding to the asymptotically Minkowski spacetime), is
consistent both in the Vainshtein regime ($r\ll r_V$) and at
large distances, provided $k_2 >0$ and $k_3\alpha \geq 0$.

The asymptotically de Sitter case has been treated separately in
Sec.~\ref{Sec6}. It should be noted that the self-accelerating de
Sitter stage is only possible for the choice $k_2 <0$ in action
(\ref{Eq1}). Therefore naively one would expect appearance of a
ghost instability at least in the Vainshtein regime, see e.g.
Eq.~(\ref{Eq20}), where a negative $k_2$ means ghost scalar
perturbation. One should also have in mind that the existence of
the Vainshtein regime is generic, even with no direct
scalar-matter coupling, because of the {\it induced} coupling for
time-dependent solutions. It turns out, however, that the choice
of the correct branch, corresponding to the asymptotically de
Sitter solution, namely, Eq.~(\ref{Eq8}) with the {\it upper}
sign (in contrast to the asymptotically Minkowski where the {\it
lower} sign is taken), changes drastically the behavior of
perturbations. Indeed, we found that the solution is stable for
the Vainshtein regime, contrary to naive expectations.

The large distance behavior, $r\gg r_V$, for the asymptotic de
Sitter is more subtle. The pure cosmological perturbations of the
spin-0 degree of freedom have positive-energy dust-like behavior
(i.e., with vanishing sound speed). Therefore any small deviation
can potentially create a change in signature for perturbations.
In fact, the deviations caused by a local material body do not
yield a correct signature: The field equations for perturbations
become elliptic in the angular directions. However, this problem
exists only in the pure de Sitter universe filled with the
Galileon field with no external matter. In a more realistic
scenario, when a small amount of normal matter is added (such
that the universe is almost de Sitter), the effective metric
(\ref{Eq16}) always has the correct signature.

Another interesting effect, associated with the cosmological time
evolution of the Galileon, is the appearance of a significant
friction term in the equation of motion. Because of the
commutation of covariant derivatives, the Galileon field
equations involve couplings of the scalar field to the curvature
tensor. In the cubic model (\ref{Eq1}) studied in this paper,
this generates \textit{within} matter an extra interaction of
scalar perturbations with their background. We showed that it
causes an efficient damping of these perturbations, a local
effect paradoxically caused by the cosmological time evolution.

To summarize, we studied spherically symmetric solutions of the
cubic covariant Galileon model in presence of a matter source and
a cosmological evolution. We found that the time evolution of the
Galileon, even as small as a cosmological one, leads to
considerable effects: in particular, the appearance of an {\it
induced} coupling, which is generically of the order of unity, so
that the Galileon becomes effectively coupled to local matter
sources even if the bare coupling is zero; and the emergence of a
friction term, which effectively {\it damps} perturbations within
matter sources. The detailed analysis of perturbations showed
that the Galileon model is well behaved in asymptotically
Minkowski space as well as in the asymptotically de Sitter,
provided that the correct branch of the solution is chosen.

\section*{Acknowledgments}
The work of E.B. was supported in part by grant FQXi-MGA-1209
{}from the Foundational Questions Institute. The work of G.E.-F.
was in part supported by the ANR grant THALES.


\end{document}